\documentclass[letterpaper, 10 pt, conference]{ieeeconf}  % Comment this line out if you need a4paper

\IEEEoverridecommandlockouts                              % This command is only needed if 
\usepackage{epsfig} % for postscript graphics files
\usepackage{cite}

\title{\LARGE \bf
Agents facilitate one category of human empathy through task difficulty
}

\author{Takahiro Tsumura$^{1}$ and Seiji Yamada$^{1}$ % <-this % stops a space
\thanks{$^{1}$Takahiro Tsumura and Seiji Yamada is with Department of Informatics, The Graduate University for Advanced Studies, SOKENDAI, Tokyo, Japan.
Also Takahiro Tsumura and Seiji Yamada is with Digital Content and Media Sciences Research Division, National Institute of Informatics, Tokyo, Japan.
        }%
}
\begin{document}

\maketitle
\thispagestyle{empty}
\pagestyle{empty}

%%%%%%%%%%%%%%%%%%%%%%%%%%%%%%%%%%%%%%%%%%%%%%%%%%%%%%%%%%%%%%%%%%%%%%%%%%%%%%%%
\begin{abstract}
One way to improve the relationship between humans and anthropomorphic agents is to have humans empathize with the agents. 
In this study, we focused on a task between agents and humans. 
We experimentally investigated hypotheses stating that task difficulty and task content facilitate human empathy. 
The experiment was a two-way analysis of variance (ANOVA) with four conditions: task difficulty (high, low) and task content (competitive, cooperative). 
The results showed no main effect for the task content factor and a significant main effect for the task difficulty factor. 
In addition, pre-task empathy toward the agent decreased after the task. The ANOVA showed that one category of empathy toward the agent increased when the task difficulty was higher than when it was lower.
This indicated that this category of empathy was more likely to be affected by the task. 
The task itself used can be an important factor when manipulating each category of empathy.
\end{abstract}

%%%%%%%%%%%%%%%%%%%%%%%%%%%%%%%%%%%%%%%%%%%%%%%%%%%%%%%%%%%%%%%%%%%%%%%%%%%%%%%%
\section{INTRODUCTION}
Humans use a variety of tools to perform tasks in society. 
Media acquisition is when humans treat artifacts like humans \cite{Reeves96}. 
It has been shown that humans develop similar feelings toward artifacts as they do toward other humans. 
In fact, there are examples of humans empathizing with artifacts in the same way that they empathize with humans. 
Typical examples of artifacts we empathize with include cleaning robots, pet-type robots, characters in competitive games, and anthropomorphic agents that provide services such as for online shopping and at help desks. 
However, there are a certain number of humans who cannot accept agents \cite{Nomura06,Nomura08,Nomura16}. 
Such agents are already in use in human society and coexist with humans. 
However, when a task is performed by a human and an agent, there is a problem when the human considers the agent to be a tool. 
When we use agents as tools, we may not need to empathize with them, but the result is that we treat agents less well when they are used in place of humans. 
Therefore, empathy from the human to the agent is essential when the agent is a supporter or competitor of the human. 
Humans and anthropomorphic agents already interact in competitive and cooperative tasks. 
For humans to have a good relationship with agents, they need to have empathy for them. 
Empathy makes it easier for humans to take positive action toward an agent and accept the agent. 
Although there have been various studies on factors that cause empathy, such as verbal and non-verbal information, situations, and relationships, we focus on tasks and experimentally examine how task difficulty and task content affect empathy. 
In this study, we focus on human-agent tasks and conduct experiments to investigate whether humans empathize more with agents depending on task difficulty and task content.

\section{RELATED WORK}
In the field of psychology, empathy has been the focus of much attention and research. 
Omdahl \cite{Omdahl95} classified empathy into three main categories: (1) affective empathy, which is an emotional response to another person's emotional state, (2) cognitive empathy, which is a cognitive understanding of another person's emotional state, and (3) empathy that includes both of the above. 
Preston and De Waal \cite{Preston02} proposed that at the heart of empathic responses is a mechanism that allows the observer access to the subjective emotional state of the subject. 
The Perception-Action Model (PAM) was defined by them to unify the differences in empathy. 
They defined empathy as a total of three types: (a) sharing or being affected by the emotional states of others, (b) evaluating the reasons for emotional states, and (c) the ability to identify and incorporate the perspectives of others. 
Olderbak et al. \cite{Olderbak14} described theoretical and empirical support for the affective specificity of empathy and developed an emotion-specific empathy questionnaire that assesses affective and cognitive empathy for six basic emotions.
\\ \indent
Various questionnaires are used as measures of empathy, but we used  the Interpersonal Reactivity Index (IRI). 
IRI, also used in the field of psychology, is used to investigate the characteristics of empathy \cite{Davis80}. 
There is another questionnaire, the Empathy Quotient (EQ)\cite{Cohen04}, but we did not use it in our study because we wanted to investigate which categories of empathy were affected after experiencing the task.
\\ \indent
In the fields of human-agent interaction (HAI), human-robot interaction (HRI), and human-computer interaction (HCI), empathy between humans and agents or robots is studied. 
In the field of HCI, research is focused on empathy. 
Wright and McCarthy \cite{Wright08} discussed the use of empathy, citing studies that have used empathy in HCI. 
Pratte et al. \cite{Pratte21} analyzed 26 publications on empathy tools and developed a framework for empathy tool designers.
\\ \indent
The following studies have been conducted in various areas of HRI. 
Leite et al. \cite{Leite14} conducted a long-term study in elementary schools to present and evaluate an empathy model for a social robot that interacts with children over a long period of time. 
They measured children's perceptions of social presence, engagement, and social support. 
Zhi et al. \cite{Zhi18} examined whether robots can have the social influence to induce nearby bystanders to stop the abuse by humans and actively defend against human abuse.
\\ \indent
In addition, the following studies have been conducted in the field of HAI. 
Richards et al. \cite{Richards18} noted that it is important to design intelligent virtual agents that influence users' emotions and intrinsic motivations. 
They determined the circumstances in which users respond to different verbal expressions of empathy. 
Okanda et al. \cite{Okanda19} focused on appearance and investigated Japanese adults' beliefs about friendship and morality toward robots. 
They examined whether the appearances of robots (i.e., humanoid, dog-like, oval-shaped) differed in relation to their animistic tendencies and empathy.
\\ \indent
A study of cooperative and competitive tasks was conducted by Ruissen and de Bruijn \cite{Ruissen16}. 
In the study, cooperative and competitive tasks were tested using Tetris. 
The results confirmed that the cooperative task did not reduce self-integration, but the competitive task did. 
Another study of competitive tasks between humans and robots is that of Kshirsagar et al. \cite{Kshirsagar19}. 
They performed a human-robot competitive task using the same task and found that participants preferred a lower-performing robot to a higher-performing one.
 Boucher et al. \cite{Boucher12} performed a human-robot cooperation task. 
The robot recognized gaze guidance to the human faster than the robot gave voice instructions to the human.
\\ \indent
Some studies of task difficulty include the following. 
Fuentes-García et al. \cite{Fuentes19} used chess problem-solving tasks of different difficulty levels to investigate participants' heart rate variability in terms of difficulty, stress, complexity, and cognitive needs. 
Cho \cite{Cho21} considered that task difficulty and mental workload are necessary to improve the usability and frequency of use of interactive systems and proposed a new approach for automatically estimating task difficulty by focusing on human blinking.
\\ \indent
Paiva defined the relationship between human beings and empathic agents, referred to as empathy agents, as designed in previous HAI and HRI research. 
As a definition of empathy between an anthropomorphic agent or robot and a human, Paiva represented empathy agents in two different ways and illustrated them \cite{Paiva04,Paiva11,Paiva17}: A) targets to be empathized with by humans and B) observers who empathize with humans.
\\ \indent
We have taken these figures and summarized them in Figure 1. 
T stands for target, E for event, and O for observer, and t represents the passage of time. 
The arrows from T and E to O are the information that O needs to empathize with targets. 
This information includes facial expressions, behaviors, and changes in the surrounding environment. 
O responds empathetically to this information and T's information as well.
This is the white arrow. 
Empathy agents and humans are divided into T and O. When the agent is T, it is an empathic target agent, and when the agent is O, it is an empathic observer agent. 
In this study, we use the empathic target agent to promote human empathy.
\begin{figure}[tbp]
    \centering
    \includegraphics[scale=0.23]{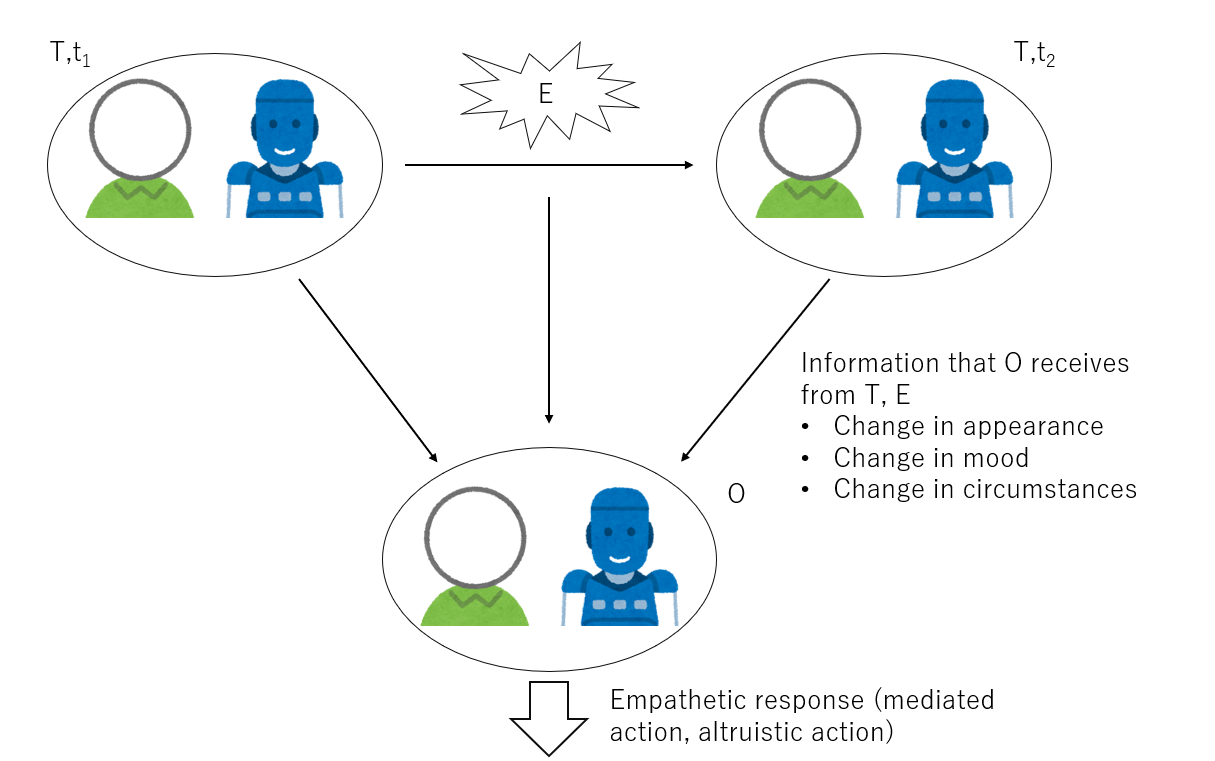}
    \caption{Conceptual diagram of empathy agent}
    \label{fig:my_label}
\end{figure}

\section{EXPERIMENTAL METHODS}
\subsection{Experimental goals and design}
The purpose of this study is to investigate whether task difficulty and task content can elicit more human empathy in an interaction with an empathy agent.
We are the first study to relate task difficulty and task content to empathy and apply it to HAI. 
This research will facilitate the application of agents used in human society by influencing human empathy. 
In addition, if there is a change in human empathy due to the influence of a task, the importance of the task can be discussed among humans. 
For these purposes, we developed two hypotheses.
\begin{itemize}
\item When performing a competitive task with an empathy agent, the higher the task difficulty, the more human empathy is suppressed.
\item When cooperating with an empathy agent, the higher the task difficulty, the more human empathy is promoted.
\end{itemize}

The above hypotheses were determined by inferring from the results of the Ruissen and de Bruijn \cite{Ruissen16} and Fuentes-García et al. \cite{Fuentes19} studies.
The above hypotheses were reached because, in cooperative tasks, humans improve their performance and have favorable impressions of their cooperating partners, whereas in competitive tasks, they think less about their adversaries. 
\\ \indent
In addition, as empathy changes with task difficulty, the higher the difficulty, the greater the mental load, and the greater the impact on performance. 
Therefore, we hypothesized that, in competitive tasks, the higher the task difficulty, the more human empathy would be suppressed. 
\\ \indent
In comparison, in the cooperative task, task performance was improved by quickly reading the intentions of the cooperating partner, which may be related to perspective taking in the cognitive empathy category. 
Since the task was facilitated by putting oneself in the other person's shoes and reading the other person's actions, we hypothesized that human empathy is facilitated in cooperative tasks as the task difficulty increases.
\\ \indent
An experiment was conducted to investigate these hypotheses with a two-factor between-participants design with two factors: task difficulty and task content. 
The number of levels for each factor was two for difficulty (high, low) and two for content (competitive, cooperative). 
Participants took part in only one of four different content conditions. 
The dependent variable was the empathy held by the participants.
\begin{figure}[tbp]
		\begin{center}
		\includegraphics[scale=0.25]{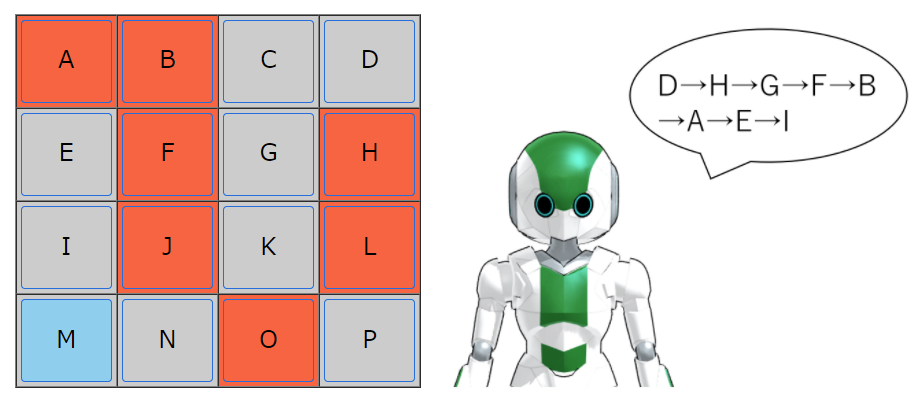}
		\caption{Task scene with empathy agent during high difficulty}
		\label{fig:taka}
	\end{center}
\end{figure}
\begin{figure}[tbp]
		\begin{center}
		\includegraphics[scale=0.25]{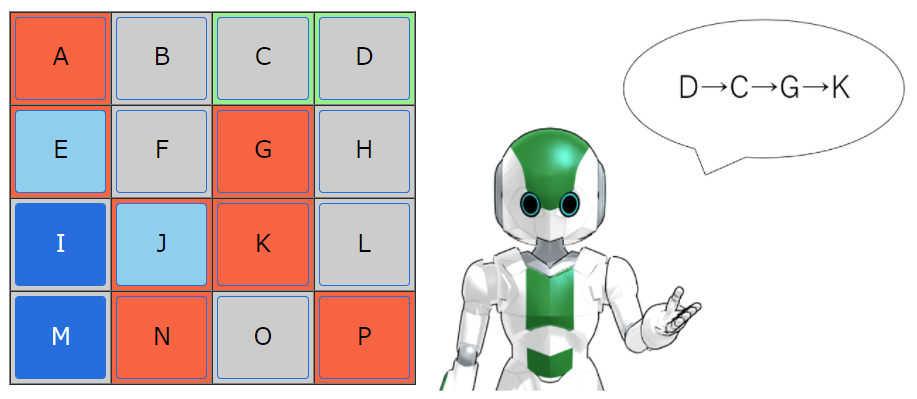}
		\caption{Task scene with empathy agent during low difficulty}
		\label{fig:robota}
	\end{center}
\end{figure}

\subsection{Experimental details}
The experiment was conducted in an online environment. 
The online environment used has already become a common method of experimentation \cite{Davis99,Crump13,Okamura20}. 
As mentioned, the purpose of this study is to promote human empathy toward anthropomorphic agents. 
When performing a task with an anthropomorphic agent, the environment is assumed to be accessed via a PC rather than in reality, so we thought that the same effect could be achieved even with an online environment.
\\ \indent
Before performing the task, a questionnaire was administered to measure the empathy toward the empathy agent. 
At this point, participants were not allowed to make judgments about the task difficulty or task content. 
The reason for administering the questionnaire before the task was to check for differences in participants' empathy toward the agent.
\\ \indent
Tasks were set up differently depending on the content of the competition and cooperation, but to avoid drastic changes in task content, the common denominator was to move between squares within a range of 16 (4 × 4 squares). 
Common to all tasks was that participants always moved from the bottom left square, and the agent always moved from the top right square. 
The mass movement were above, below, to the left, and to the right of the current point, and the same square could only be passed through once. 
The total number of moves that could be made alternated, but the total number for the agent was indicated at the start. 
The reason why this information for the agent was given advance is that it increased the difficulty of the task and reduced the difference in difficulty between the two difficulty levels. 
If this information were not given, participants would need to think about how many moves the agent could make, which would be burdensome and would affect the judgment of the difficulty level. 
The task was made simple so that the comparison of difficulty levels would not be affected by the selections of the agent while trying to anticipate their movement total. 
Every task had a checkpoint, and the purpose was to pass through this point. 
The squares selected by the participants were colored blue, the squares that could be moved to next were colored light blue, the squares selected by the agent were colored light green, and the checkpoints were colored vermilion.
\\ \indent
The task was performed three times in total, with different checkpoint locations. 
The total number of moves varied depending on the difficulty level of the task: eight for the high difficulty and four for the low difficulty. 
In the competitive task, the purpose was to pass more checkpoints than the opponent, while in the cooperative task, the purpose was to pass checkpoints through cooperation. 
Other detailed conditions are explained in later sections.
\\ \indent
To examine only the factors of interest in this experiment, the appearance and behavior of the agents were also standardized. 
In addition, agents did not speak during the task but only made minimal gestures. 
Participants interacted with an agent in one of four conditions, combining task difficulty (high, low) and task content (competitive, cooperative).
\\ \indent
Afterwards, the participants' empathy toward the agent was aggregated by another questionnaire similar to before the task. 
Then, the participants were asked to write their impressions of the experiment in free form.

\subsection{Factor detail setting}
\subsubsection{Task difficulty}
Two task difficulty levels were prepared for this experiment. 
Figures 2 and 3 show these levels. 
The following conditions were used for different levels of difficulty.
\\ \indent
A) The total number of squares that could be moved to was eight for the high difficulty level and four for the low difficulty level. 
B) The number of checkpoints was seven for the high difficulty level and five for the low difficulty level. 
C) In the case of the competitive task, the high difficulty level required the participant to act in such a way that at least four checkpoints were passed, while the low difficulty level required them to act in such a way that at least two checkpoints were passed. 
D) In the case of the cooperative task, for the high difficulty level, the human and agent had to cooperate to pass all seven checkpoints, and for the low difficulty level, they had to cooperate to pass at least four checkpoints.
\\ \indent
By reducing the total number of moves to be made by half and simplifying the expected number of checkpoints that the participants and agents had to pass through for the low-difficulty level, the number of trials was made to have a significant effect on the difficulty level.

\subsubsection{Task content} 
Two types of task content were prepared. 
By keeping the task environments as close as possible, we eliminated external factors and tried to measure the effect of task content on human empathy. 
The two types of task content were competitive and cooperative.
\\ \indent
In the competitive task, the task was a checkpoint competition, and the number of checkpoints required to win varied depending on the difficulty level. 
Points were awarded to the first person to pass each checkpoint. 
The win ratio for a total of three tasks was one win, one loss, and one tie, even when participants took the optimal actions. 
The win rate was adjusted to reduce the impact of the win rate on human empathy.
\\ \indent
In the cooperative task, the purpose of the task was for the participant and agent to pass all the checkpoints, and the total number of checkpoints varied with the difficulty level. 
The high difficulty level required the human and agent to cooperate to pass all seven checkpoints, and the low difficulty level required them to cooperate to pass at least four checkpoints. 
The maximum number of checkpoints that can be passed by each participant in a total of three tasks was adjusted. 
This was done to prevent one side from always passing too many odd-numbered checkpoints. 
It does not make sense if both parties pass through the same checkpoint.

\subsection{Questionnaire}
Participants completed a questionnaire before and after the task. 
The questionnaire was a 12-item questionnaire modified from the Interpersonal Reactivity Index (IRI), which is used to investigate the characteristics of empathy, to suit the present experiment \cite{Davis80}. 
The two questionnaires were the same. 
Both used were based on the IRI and were surveyed on a 5-point Likert scale (1: not applicable, 5: applicable). 
The questionnaire used is shown in Table 1. Q4, Q9, and Q10 are inverted items, so the scores were reversed when analyzing them.
\begin{table*}[tbp] 
    \caption{Summary of questionnaire used in this experiment}
    \centering
    \scalebox{0.95}{
    \begin{tabular}{|c|c|c|l|}\hline
        \multicolumn{3}{|c|}{Empathy} & Contents \\ \hline
        & & Q1 & If an emergency happens to the character, you would be anxious and restless. \\ \cline{3-4}
        & Personal & Q2 & If the character is emotionally disturbed, you would not know what to do. \\ \cline{3-4}
        Affective & distress & Q3 & If you see the character in need of immediate help, you would be confused and would not know what to do. \\ \cline{2-4}
        empathy & & Q4 & If you see the character in trouble, you would not feel sorry for that character. \\ \cline{3-4}
        & Empathic & Q5 & If you see the character being taken advantage of by others, you would feel like you want to protect that character. \\ \cline{3-4}
        & concern & Q6 & The character's story and the events that have taken place move you strongly. \\ \hline
        & & Q7 & You look at both the character's position and the human position. \\ \cline{3-4}
        & Perspective & Q8 & If you were trying to get to know the character better, you would imagine how that character sees things.\\ \cline{3-4}
        Cognitive & taking & Q9 & When you think you're right, you don't listen to what the character has to say. \\ \cline{2-4}
        empathy & & Q10 & You are objective without being drawn into the character's story or the events taken place. \\ \cline{3-4}
        & Fantasy & Q11 & You imagine how you would feel if the events that happened to the character happened to you. \\ \cline{3-4}
        & scale & Q12 & You get deep into the feelings of the character. \\ \hline
    \end{tabular}}
    \label{tab:my_label}
\end{table*}

\subsection{Analysis method}
The analysis was a two-factor analysis of variance (ANOVA). 
The between-participant factors were two levels of task difficulty and two levels of task content.
On the basis of the results of the participants' questionnaires, we investigated how task difficulty and task content influenced the promotion of empathy as factors that elicit human empathy. 
The numerical values of empathy aggregated before and after the task were used as the dependent variable.
R (R ver. 4.1.0) was used for the ANOVA. 
Also, we used anovakun (ver. 4.8.6) as the R package.

\section{EXPERIMENTAL RESULTS}
\subsection{Experimental environment}
Participants were recruited for the experiment using the Yahoo! crowdsourcing company. 
Participants were paid 55 yen after completing all tasks as a reward for participating. 
A website was created for the experiment, which was limited to using a PC.

\subsection{Participants}
There were a total of 596 participants.
However, there were 18 participants who gave inappropriate responses, which were eliminated as erroneous data, leaving a total of 578 participants. 
To judge whether answers were inappropriate in the experiment, we judged answers as inappropriate when the changes in the empathy values before and after the video were the same for all items or when only one item changed \cite{Schonlau15,Leiner19}. 
The task of aligning the number of participants to an appropriate number for the analysis was performed, and 142 participants in each condition were included in the analysis, starting from the top in the order of their participation in the experiment. 
Thus, the total number of participants used in the analysis was 568.
\\ \indent
The average age was 48.32 years (standard deviation 11.02), with a minimum of 18 years and a maximum of 87 years. 
The gender breakdown was 344 males and 224 females.

\subsection{Analysis Result}
All 12 questionnaire items were analyzed together. 
We also categorized and analyzed affective and cognitive empathy. 
For multiple comparisons, Holm's multiple comparison test was used to examine the existence of significant differences. 
Table 2 shows the results of the overall analysis. 
The results of the questionnaire analysis showed a main effect of task difficulty on affective empathy. 
The results are shown in Figure 4.
\begin{table*}[tbp]
    \caption{Results of all analyses of variance}
    \scalebox{1.2}{
        \begin{tabular}{c|c|c||c|c||c||c|c|c}\hline 
        \multicolumn{2}{c|}{Category} & Conditions & Mean & S.D. & Factor & \em{F} & \em{p} & $\eta^2_p$\\ \hline 
        & & high-competitive & 38.7746 & 6.0675 & difficulty & 0.7877 & 0.3752 \em{ns} & 0.0014 \\ \cline{3-9}
        & & high-cooperative & 38.4366 & 5.6949 & content & 0.6676 & 0.4142 \em{ns} &  0.0012 \\ \cline{3-9}
        Empathy & pre & low-competitive & 38.4014 & 6.1009 & interaction & 0.0198 & 0.8880 \em{ns} & 0.0000 \\ \cline{3-9}
        & & low-cooperative & 37.9225 & 5.9562 & & & & \\ \cline{2-9}
        & & high-competitive & 37.7535 & 6.5852 & difficulty & 2.4737 & 0.1163 \em{ns} & 0.0044 \\ \cline{3-9}
        (Q1-Q12) & & high-cooperative & 37.8169 & 6.0980 & content & 0.1053 & 0.7456 \em{ns} & 0.0002 \\ \cline{3-9}
        & post & low-competitive & 37.1127 & 7.3389 & interaction & 0.1909 & 0.6624 \em{ns} & 0.0003 \\ \cline{3-9}
        & & low-cooperative & 36.6831 & 6.8098 & & & & \\ \hline\hline
        & & high-competitive & 19.4859 & 3.6021 & difficulty & 2.2672 & 0.1327 \em{ns} & 0.0040  \\ \cline{3-9}
        & & high-cooperative & 19.5352 & 3.6059 & content & 0.1374 & 0.7110 \em{ns} & 0.0002 \\ \cline{3-9}
        Affective & pre & low-competitive & 19.1901 & 3.7282 & interaction & 0.2839 & 0.5944 \em{ns} & 0.0005 \\ \cline{3-9}
        empathy & & low-cooperative & 18.9155 & 3.5520 & & & & \\ \cline{2-9}
        & & high-competitive & 18.8662 & 3.8798 & difficulty & 4.0986 & 0.0434 * & 0.0072 \\ \cline{3-9}
        (Q1-Q6) & & high-cooperative & 18.9648 & 3.5479 & content & 0.1855 & 0.6668 \em{ns} & 0.0003\\ \cline{3-9}
        & post & low-competitive & 18.4437 & 4.1642 & interaction & 0.5362 & 0.4643 \em{ns} & 0.0009 \\ \cline{3-9}
        & & low-cooperative & 18.0634 & 3.9683 & & & & \\ \hline\hline
        & & high-competitive & 19.2887 & 3.2586 & difficulty & 0.0027 & 0.9584 \em{ns} & 0.0000\\ \cline{3-9}
        & & high-cooperative & 18.9014 & 3.0325 & content & 1.2025 & 0.2733 \em{ns} & 0.0021 \\ \cline{3-9}
        Cognitive & pre & low-competitive & 19.2113 & 3.2109 & interaction & 0.1152 & 0.7344 \em{ns} & 0.0002 \\ \cline{3-9}
        empathy & & low-cooperative & 19.0070 & 3.3464 & & & & \\ \cline{2-9}
        & & high-competitive & 18.8873 & 3.2900 & difficulty & 0.5918 & 0.4421 \em{ns} & 0.0010 \\ \cline{3-9}
        (Q7-Q12) & & high-cooperative & 18.8521 & 3.4124 & content & 0.0208 & 0.8854 \em{ns} & 0.0000 \\ \cline{3-9}
        & post & low-competitive & 18.6690 & 3.7088 & interaction & 0.0006 & 0.9808 \em{ns} & 0.0000 \\ \cline{3-9}
        & & low-cooperative & 18.6197 & 3.5385 & & & & \\ \hline
        \end{tabular}} \\
    \em{p}:
{{*}p\textless\em{0.05}}
{{**}p\textless\em{0.01}}
{{***}p\textless\em{0.001}}
    \label{table2}
\end{table*}
\begin{figure}[tbp]
    \centering
    \includegraphics[scale=0.25]{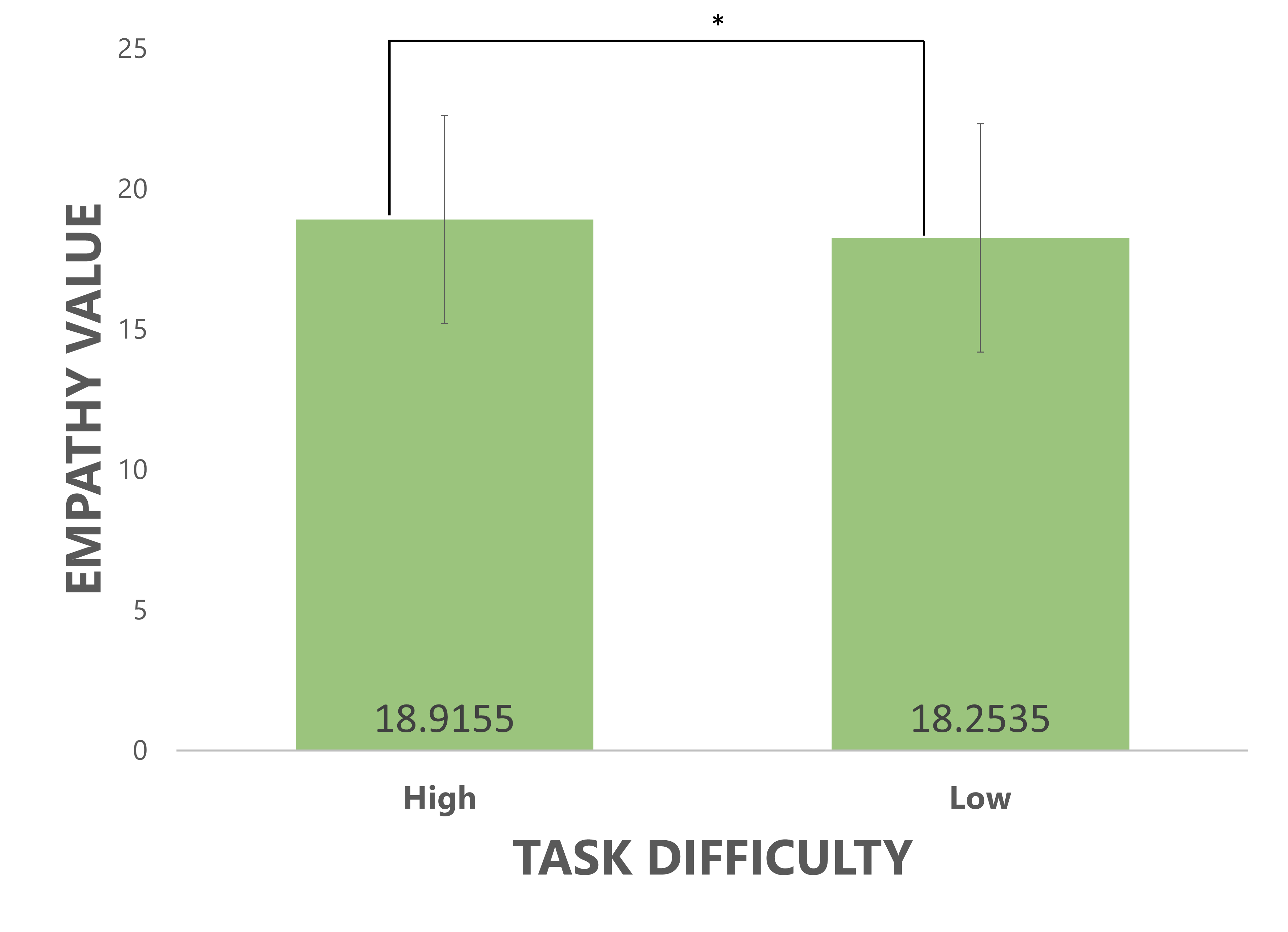}
    \caption{Main effects results of affective empathy}
    \label{fig:my_label2}
\end{figure}
\\ \indent
Initially, as can be seen from Table 2, we examined differences in empathy toward the agent among the participants and found no differences between any of the conditions. 
Therefore, we assumed that the ability to empathize with the agent was similar among the participants.
In this study, the questionnaire on pre-task empathy toward the agent was given only to confirm that there were no differences between participants. 
Therefore, we do not discuss significant differences between the pre- and post-task cases.
\\ \indent
The post-task results showed no interaction between task difficulty and task content, regardless of empathy category. 
For the 12 items, there was also no main effect of each factor [$F$(1,564) = 2.4737]. 
Similarly, no main effect of task content was found [$F$(1,564) = 0.5918]. However, a main effect was found for task difficulty [$F$(1,564) = 4.0986] based on the analysis of affective empathy in Table 2.
The main effect of post-task task-difficulty was higher for affective empathy for the higher difficulty level (high difficulty: mean = 18.9155, S.D. = 3.7113; low difficulty: mean = 18.2535, S.D. = 4.0647), as shown in Figure 4. 
On the basis of the above analysis, the results of this experiment suggest that a higher task difficulty promotes affective empathy.
\\ \indent
The post-task values were lower than the pre-task values of empathy toward the agent in each condition (all pre-task: mean = 38.3838, S.D. = 5.9490; all post-task: mean = 37.3415, S.D. = 6.7214). 
Also, for affective empathy, pre-task emotional empathy values were higher (all pre-task: mean = 19.2817, S.D. = 3.6216; all post-task: mean = 18.5845, S.D. = 3.9027). 
Similarly, for cognitive empathy, pre-task cognitive empathy values were higher (all pre-task: mean = 19.1021, S.D. = 3.2094; all post-task: mean = 18.7570, S.D. = 3.4835). 
\\ \indent
Since a main effect was found for task difficulty in affective empathy, we conducted an analysis by category of affective empathy.
The results showed that only personal distress had a main effect on task difficulty [$F$(1,564) = 5.2007].
The main effect of post-task task-difficulty was higher for affective empathy for the higher difficulty level (high difficulty: mean = 9.2148, S.D. = 2.5428; low difficulty: mean = 8.7148, S.D. = 2.6769), as shown in Figure 5. 
\begin{table*}[tbp]
    \caption{Results of affective empathy analyses of variance}
    \scalebox{1.2}{
        \begin{tabular}{c|c|c||c|c||c||c|c|c}\hline 
        \multicolumn{2}{c|}{Category} & Conditions & Mean & S.D. & Factor & \em{F} & \em{p} & $\eta^2_p$\\ \hline 
        & & high-competitive & 9.5211 & 2.6004 & difficulty & 2.5238 & 0.1127 \em{ns} & 0.0045 \\ \cline{3-9}
        & & high-cooperative & 9.6197 & 2.6836 & content & 0.0930 & 0.7606 \em{ns} & 0.0002 \\ \cline{3-9}
        Personal & pre & low-competitive & 9.3380 & 2.5817 & interaction & 0.5688 & 0.4510 \em{ns} & 0.0010 \\ \cline{3-9}
        distress & & low-cooperative & 9.1056 & 2.5922 & & & & \\ \cline{2-9}
        & & high-competitive & 9.2394 & 2.5263 & difficulty & 5.2007 & 0.0229 * & 0.0091 \\ \cline{3-9}
        (Q1-Q3) & & high-cooperative & 9.1901 & 2.5680 & content & 0.7521 & 0.3862 \em{ns} & 0.0013 \\ \cline{3-9}
        & post & low-competitive & 8.8803 & 2.6803 & interaction & 0.4127 & 0.5209 \em{ns} & 0.0007 \\ \cline{3-9}
        & & low-cooperative & 8.5493 & 2.6726 & & & & \\ \hline\hline
        & & high-competitive & 9.9648 & 1.8922 & difficulty & 0.4410 & 0.5069 \em{ns} & 0.0008  \\ \cline{3-9}
        & & high-cooperative & 9.9155 & 1.8850 & content & 0.0776 & 0.7807 \em{ns} & 0.0001 \\ \cline{3-9}
        Empathic & pre & low-competitive & 9.8521 & 2.0245 & interaction & 0.0005 & 0.9829 \em{ns} & 0.0000 \\ \cline{3-9}
        concern & & low-cooperative & 9.8099 & 2.0279 & & & & \\ \cline{2-9}
        & & high-competitive & 9.6268 & 2.1688 & difficulty & 0.8457 & 0.3582 \em{ns} & 0.0015 \\ \cline{3-9}
        (Q4-Q6) & & high-cooperative & 9.7746 & 1.8884 & content & 0.0783 & 0.7797 \em{ns} & 0.0001\\ \cline{3-9}
        & post & low-competitive & 9.5634 & 2.1683 & interaction & 0.3133 & 0.5759 \em{ns} & 0.0006 \\ \cline{3-9}
        & & low-cooperative & 9.5141 & 2.1561 & & & & \\ \hline
        \end{tabular}} \\
    \em{p}:
{{*}p\textless\em{0.05}}
{{**}p\textless\em{0.01}}
{{***}p\textless\em{0.001}}
    \label{table3}
\end{table*}
\begin{figure}[tbp]
    \centering
    \includegraphics[scale=0.25]{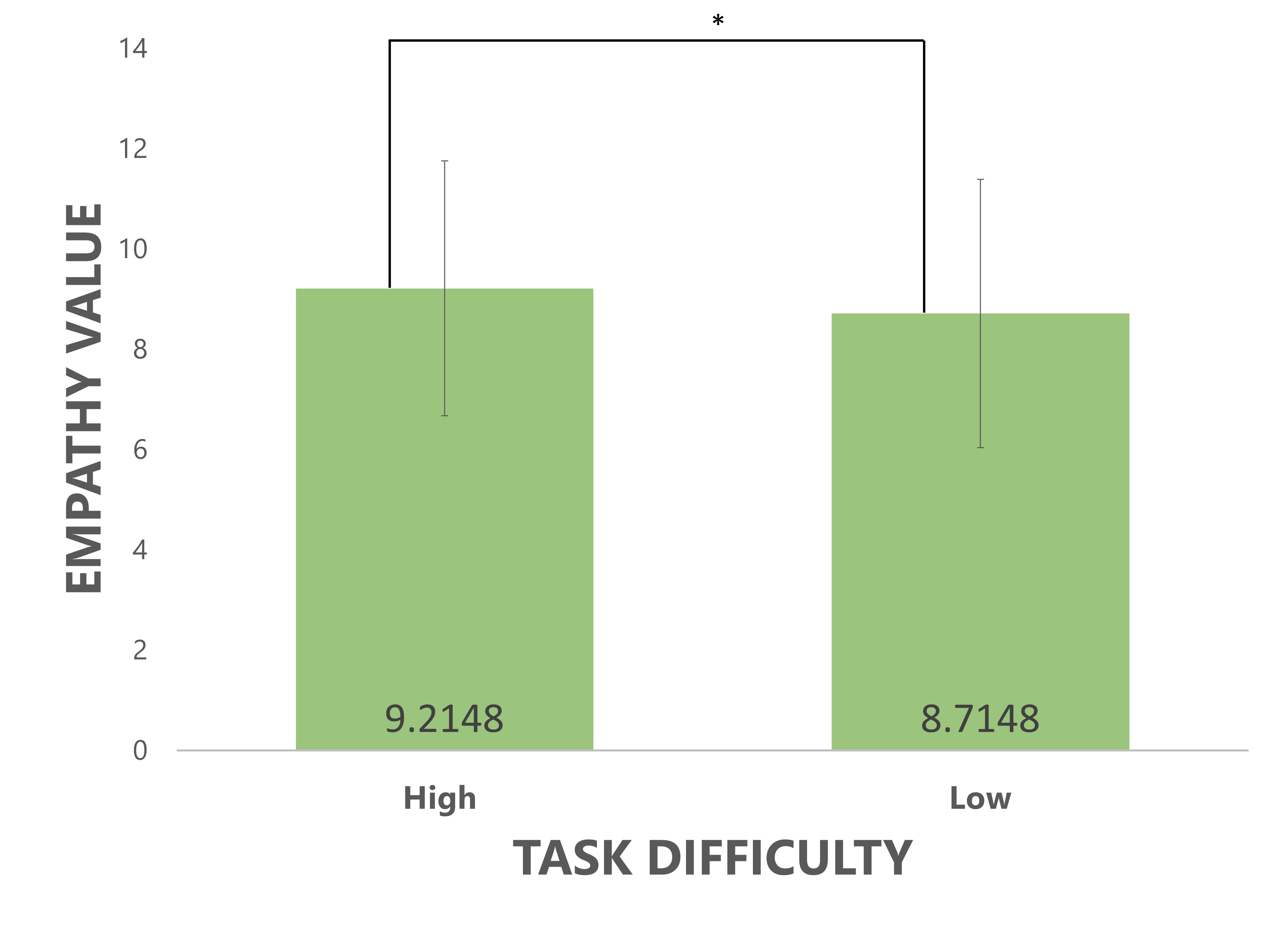}
    \caption{Main effects results of personal distress}
    \label{fig:my_label3}
\end{figure}

\section{DISCUSSION}
This experiment was conducted to investigate the conditions necessary for a human to develop empathy for an anthropomorphic agent. 
In particular, this experiment aimed to identify factors that influence the empathy between an agent and a person who performed a task by investigating factors related to the task. 
For this purpose, we formulated the following two hypotheses and analyzed the data obtained from the experiment. 
\\ \indent
The results did not support the two hypotheses. 
However, the results showed that regardless of task content, a higher task difficulty promoted human affective empathy. 
In discussing these results, it is necessary to focus on the changes in empathy before and after the task. 
\\ \indent
If only post-task surveys had been conducted, a higher task difficulty would have been shown to increase human empathy. 
However, by conducting a pre-task survey, it was found that empathy actually decreased throughout the task. 
This result indicated that post-task changes do not necessarily lead to better results than pre-task changes.
\\ \indent
In addition, the fact that empathy for the agent decreased with the task in this experiment indicates that the task may decrease empathy. 
However, as a limitation of this experiment, it is possible that the simplicity of the task itself may have decreased empathy because the task itself was perceived as tedious, due to eliminate factors other than task difficulty and task content.
\\ \indent
However, the fact that task difficulty did not affect the task content but affected human affective empathy may be an effective factor in controlling human empathy when empathic agents coexist in human society in the future. 
By setting the task difficulty appropriately, it is possible to maintain an appropriate distance from an agent without making the participant empathize more than necessary.
\\ \indent
The results of the analysis, which classified empathy into affective empathy and cognitive empathy, showed a main effect of task difficulty on affective empathy. 
The main reason for this main effect is thought to be that affective empathy increased as task difficulty increased due to the increased mental load caused by the task. 
This is related to personal distress, which is classified as affective empathy. 
Affective empathy is the feeling of the emotional state that others are experiencing or about to experience, which leads to the same emotional state in oneself. 
Only affective empathy was enhanced because the mental load from the task affected the emotional states of both the participant and agent. 
No main effects of task difficulty or task content were observed for cognitive empathy. 
This is because cognitive empathy requires imagining the thoughts and feelings of others in terms of oneself and imagining them from the other's point of view, so the task in this experiment did not enhance cognitive empathy.

\section{CONCLUSIONS}
This study focused on human-agent tasks as part of the factors and conditions that make humans empathize with anthropomorphic agents, and it investigated task difficulty and task content. 
Two hypotheses were formulated and tested. 
The results did not support either of the two hypotheses. 
Task difficulty was found to have a significant effect on affective empathy. 
The analysis revealed that a higher task difficulty increased emotional empathy after the task. 
The task itself can be an important factor when manipulating each category of empathy.
Future research may consider the development of an agent that empathizes with humans and that is suitable for a task since it was confirmed that empathy held by humans decreases.

\addtolength{\textheight}{0cm}   % This command serves to balance the column lengths

%%%%%%%%%%%%%%%%%%%%%%%%%%%%%%%%%%%%%%%%%%%%%%%%%%%%%%%%%%%%%%%%%%%%%%%%%%%%%%%%

%%%%%%%%%%%%%%%%%%%%%%%%%%%%%%%%%%%%%%%%%%%%%%%%%%%%%%%%%%%%%%%%%%%%%%%%%%%%%%%%

%%%%%%%%%%%%%%%%%%%%%%%%%%%%%%%%%%%%%%%%%%%%%%%%%%%%%%%%%%%%%%%%%%%%%%%%%%%%%%%%

%%%%%%%%%%%%%%%%%%%%%%%%%%%%%%%%%%%%%%%%%%%%%%%%%%%%%%%%%%%%%%%%%%%%%%%%%%%%%%%%

\bibliographystyle{IEEEtran}
\bibliography{test}

\end{document}